\documentclass[prd,preprint,a4paper,superscriptaddress,nofootinbib,showpacs,11pt]{revtex4}
\def\beqa{\begin{eqnarray}}
\def\eeqa{\end{eqnarray}}
\def\beq{\begin{equation}}
\def\eeq{\end{equation}}
\def\ad{\dot{a}}
\def\vol{\int d^4x\,\sqrt{-g}}

\def\half{\frac{1}{2}}
\def\gu{g^{\mu\nu}}
\def\gd{g_{\mu\nu}}
\def\umu{^{\mu}}

\def\umunu{^{\mu\nu}}
\def\dmunu{_{\mu\nu}}
\def\ua{^{\alpha}}

\def\da{_{\alpha}}

\def\dab{_{\alpha\beta}}

\def\ddeab{_{;\alpha\beta}}
\def\ddemunu{_{;\mu\nu}}

\def\ddemu{_{;\mu}}  
\def\ddenu{_{;\nu}}  
\def\ddea{_{;\alpha}}  \def\udea{^{;\alpha}}
\def\ddeb{_{;\beta}}

\def\pa{\partial}

\let\lam=\lambda

\let\gam=\gamma
\let\alp=\alpha

\renewcommand{\epsilon}{\varepsilon}

\def\bet{\begin{tabular}}
\def\eet{\end{tabular}}
\def\bef{\begin{figure}}
\def\eef{\end{figure}}
\def\beqa{\begin{eqnarray}}
\def\eeqa{\end{eqnarray}}
\def\beq{\begin{equation}}
\def\eeq{\end{equation}}
\def\vol{\int d^4x\,\sqrt{-g}}

\def\half{\frac{1}{2}}

\def\gu{g^{\mu\nu}}
\def\gd{g_{\mu\nu}}

\def\umu{^{\mu}}

\def\umunu{^{\mu\nu}}
\def\dmunu{_{\mu\nu}}
\def\ua{^{\alpha}}

\def\da{_{\alpha}}

\def\dab{_{\alpha\beta}}

\def\ddeab{_{;\alpha\beta}}
\def\ddemunu{_{;\mu\nu}}

\def\ddemu{_{;\mu}}

\def\ddenu{_{;\nu}}

\def\ddea{_{;\alpha}}
\def\udea{^{;\alpha}}
\def\ddeb{_{;\beta}}

\def\pa{\partial}

\def\ie{{\it i.e. }}

\def\fp{F'(\phi)}

\def\p{\phi}

\def\l{\cal L}

\let\lam=\lambda

\let\gam=\gamma
\let\alp=\alpha

\renewcommand{\epsilon}{\varepsilon}

\def\beqa{\begin{eqnarray}}
\def\eeqa{\end{eqnarray}}
\def\beq{\begin{equation}}
\def\eeq{\end{equation}}
\def\disp{\displaystyle}

\def\vol{d^4x\,\sqrt{-g}}

\def\half{\frac{1}{2}}
\def\gu{g^{\mu\nu}}
\def\gd{g_{\mu\nu}}
\def\gbru{{\bar{g}}^{\mu\nu}}
\def\gbrd{{\bar{g}}_{\mu\nu}}

\def\pa{\partial}

\def\al{{\alpha}}

\def\gam{{\gamma}}

\def\lam{{\lambda}}

\def\ome{{\omega}}
\def\GAM{{\Gamma}}
\def\LAM{{\Lambda}}

\def\ua{^{\alpha}}

\def\da{_{\alpha}}

\def\umu{^{\mu}}

\def\umunu{^{\mu\nu}}
\def\dmunu{_{\mu\nu}}

\def\dab{_{\alpha\beta}}

\def\udea{_{;}^{\alpha}}

\def\ddemu{_{;\mu}}
\def\ddenu{_{;\nu}}
\def\ddea{_{;\alpha}}
\def\ddeb{_{;\beta}}
\def\ddemunu{_{;\mu\nu}}

\def\ddeab{_{;\alpha\beta}}

\def\f{{\phi}}

\def\gbr{{\bar{g}}}
\def\ad{{\dot{a}}}
\def\add{{\ddot{a}}}
\def\ap{{a^{'}}}

\def\abr{{\bar{a}}}
\def\abrd{{\dot{\bar{a}}}}
\def\abrdd{{\ddot{\bar{a}}}}

\def\abrpq{{{\bar{a}}^{'2}}}
\def\fd{{\dot{\phi}}}
\def\fdd{{\ddot{\phi}}}
\def\fp{{\phi^{'}}}

\def\fpq{{\phi^{'2}}}
\def\fbr{{\bar{\phi}}}
\def\fbrd{{\dot{\bar{\phi}}}}
\def\fbrdd{{\ddot{\bar{\phi}}}}

\def\fbrpq{{{\bar{\phi}}^{'2}}}
\def\VDEF{{V_{\phi}}}
\def\VBR{{\bar{V}}}
\def\VBRDEF{{\bar{V}_{\bar{\phi}}}}
\def\FDEF{{F_{\phi}}}
\def\FD{{\dot{F}}}

\def\tbr{{\bar{t}}}
\def\LBR{{\overline{L}}}
\def\EBR{{\overline{E}}}

\def\ie{{\it i.e. }}

\def\pr{{\it Phys. Rev.}\ }
\def\prl{{\it Phys. Rev. Lett.}\ }
\def\pl{{\it Phys. Lett.}\ }
\def\np{{\it Nucl. Phys.}\ }

\def\ijmp{{\it Int. Journ. Mod. Phys.}\ }
\def\ijtp{{\it Int. Journ. Theor. Phys.}\ }
\def\aph{{\it Ann. Phys. (N.Y.)}\ }

\def\cqg{{\it Class. Quantum Grav.}\ }

\def\grg{{\it Gen. Relativ. Grav.}\ }

\def\apj{{\it Ap. J.}\ }

\def\ncim{{\it Nuovo Cim.}\ }

\def\mnras{{\it Mon. Not. R. Ast. Soc.}\ }
\def\prep{{\it Phys. Rep.}\ }

\def\araa{{\it Ann. Rev. Astr. Ap.}\ }

\def\hpa{{\it Helv. Phys. Acta}\ }

\let\lam=\lambda  

\let\gam=\gamma
\let\alp=\alpha

\renewcommand{\epsilon}{\varepsilon}

\def\pr{{\it Phys. Rev.}\ }
\def\prl{{\it Phys. Rev. Lett.}\ }
\def\pl{{\it Phys. Lett.}\ }
\def\np{{\it Nucl. Phys.}\ }

\def\ijmp{{\it Int. Journ. Mod. Phys.}\ }
\def\ijtp{{\it Int. Journ. Theor. Phys.}\ }

\def\cqg{{\it Class. Quantum Grav.}\ }

\def\grg{{\it Gen. Relativ. Grav.}\ }

\def\apj{{\it Ap. J.}\ }

\def\ncim{{\it Nuovo Cim.}\ }

\def\mnras{{\it Mon. Not. R. Ast. Soc.}\ }
\def\prep{{\it Phys. Rep.}\ }

\def\araa{{\it Ann. Rev. Astr. Ap.}\ }

\def\ie{{\it i.e. }}

\def\fp{F'(\phi)}

\def\p{\phi}

\def\l{\cal L}

\usepackage{amsfonts}
\usepackage{amssymb}
\usepackage{hyperref}
\usepackage{graphics}
\def\fs{\footnotesize}

\begin{document}

\title{\Large Conformal aspects of Palatini approach in Extended Theories of Gravity}

\date{\today}

\author{Gianluca Allemandi}
\email{allemandi@dm.unito.it} \affiliation{{\fs Dipartimento di
Matematica}, {\fs Universit\`a di Torino}\\{\fs Via C. Alberto 10,
10123 TORINO (Italy)}}

\author{Monica Capone}
\email{capone@sa.infn.it}
\affiliation{{\fs Dipartimento di Fisica
``E.R. Caianiello"}, {\fs Universit\`a di Salerno and INFN. Sez. di
Napoli}\\{\fs Via S. Allende, I-84081 Baronissi (SA) (Italy)}}

\author{Salvatore Capozziello}
\email{capozziello@sa.infn.it}
\affiliation{{\fs Dipartimento di
Fisica ``E.R. Caianiello"}, {\fs Universit\`a di Salerno and INFN.
Sez. di Napoli}\\{\fs Via S. Allende, I-84081 Baronissi (SA)
(Italy)}}

\author{Mauro Francaviglia}
\email{francaviglia@dm.unito.it} \affiliation{{\fs Dipartimento di
Matematica}, {\fs Universit\`a di Torino}\\{\fs Via C. Alberto 10,
10123 TORINO (Italy)}}


\begin{abstract}
The debate on the physical relevance of conformal transformations
can be faced by taking the Palatini approach into account to
gravitational theories. We show that conformal transformations are
not only a mathematical tool to disentangle gravitational and
matter degrees of freedom (passing from the Jordan frame to the
Einstein frame) but they acquire a physical meaning considering
the bi-metric structure of Palatini approach which allows to
distinguish between spacetime structure and geodesic structure.
Examples of higher-order and non-minimally coupled theories are
worked out and relevant cosmological solutions in Einstein frame
and Jordan frames are discussed showing that also the
interpretation of cosmological observations can drastically change
depending on the adopted frame.

\end{abstract}

\maketitle

\section{Introduction}
Einstein's General Relativity (GR) can be considered as one of the
major scientific achievements of last century. For the first time,
a comprehensive theory of spacetime, gravity and matter has been
formulated giving rise to a new conception of the Universe.
However, in the last thirty years, several shortcomings came out
in the Einstein scheme and people  began to investigate whether GR
is the only fundamental theory  capable of explaining the
gravitational interaction. Such issues come, essentially, from
cosmology and quantum field theory. In the first case, the
presence of the Big Bang singularity, flatness and horizon
problems \cite{guth} led to the statement that Standard
Cosmological Model \cite{weinberg}, based on GR and Standard Model
of particle physics, is inadequate to describe the Universe at
extreme regimes. On the other hand, GR is a {\it classical} theory
which does not work as a fundamental theory, when one wants to
achieve a full quantum description of spacetime (and then of
gravity). Due to this facts and, first of all, to the lack of a
definitive quantum gravity theory, alternative theories of gravity
have been pursued in order to attempt, at least, a semi-classical
scheme where GR and its positive results could be recovered. One
of the most fruitful approaches has been that of {\it Extended
Theories of Gravity} (ETG)  which have become a sort of paradigm
in the study of gravitational interaction based on   corrections
and enlargements of the traditional Einstein scheme. The paradigm
consists, essentially,  in adding higher-order curvature
invariants and minimally or non-minimally coupled scalar fields
into dynamics which come out from  the effective action of quantum
gravity \cite{odintsov}. Other motivations to modify GR come from
the issue of a whole recovering of Mach principle \cite{brans}
which leads to assume a varying gravitational coupling. This
principle states that the local inertial frame is determined by
some average of the motion of distant astronomical objects
\cite{bondi}, so that gravitational coupling can be
scale-dependent and related to some scalar field. As a
consequence,  the concept of ``inertia'' and equivalence principle
have to be revised. For example, the Brans--Dicke theory is a
serious attempt to define an alternative theory to the Einstein
gravity: it takes into account a variable Newton gravitational
constant, whose dynamics is governed by a scalar field
non-minimally coupled with geometry. In such a way, Mach's
principle is better implemented \cite{brans,cimento,sciama}.

All these approaches are not the ``{\it full quantum gravity}" but
are needed as working schemes toward it. In  any case, they are
going to furnish consistent and physically reliable results.
Furthermore, every unification scheme as Superstrings,
Supergravity or Grand Unified Theories, takes into account
effective actions where non-minimal couplings to the geometry or
higher--order terms in the curvature invariants come out. Such
contributions are due to one--loop or higher--loop corrections in
the high--curvature regimes near the full (not yet available)
quantum gravity regime \cite{odintsov}. Specifically, this scheme
was adopted in order to deal with the quantization on curved
spacetimes and the result was that the interactions among quantum
scalar fields and background geometry or the gravitational
self--interactions yield corrective terms in the Einstein--Hilbert
Lagrangian \cite{birrell}. Moreover, it has been realized that
such corrective terms are inescapable if we want to obtain the
effective action of quantum gravity on scales closed to the Planck
length \cite{vilkovisky}. Higher--order terms in curvature
invariants (such as $R^{2}$, $R\umunu R\dmunu$,
$R^{\mu\nu\alp\beta}R_{\mu\nu\alp\beta}$, $R \,\Box R$, or $R
\,\Box^{k}R$) or non-minimally coupled terms between scalar fields
and geometry (such as $\p^{2}R$) have to be added to the effective
Lagrangian of gravitational field when quantum corrections are
considered. For instance, one can notice that such terms occur in
the effective Lagrangian of strings or in Kaluza--Klein theories,
when the mechanism of dimensional reduction is used
\cite{veneziano}.

 From a conceptual point of view, there would be no {\it a priori}
reason to restrict the gravitational Lagrangian to a linear
function of the Ricci scalar $R$, minimally coupled with matter
\cite{francaviglia}. Furthermore, the idea that there are no
``exact'' laws of physics but that the Lagrangians of physical
interactions are ``stochastic'' functions -- with the property
that local gauge invariances (\ie conservation laws) are well
approximated in the low energy limit and that physical constants
can vary -- has been taken into serious consideration -- see
Ref.~\cite{ottewill}.

Besides fundamental physics motivations, all these theories have
acquired a huge interest in cosmology due to the fact that they
``naturally" exhibit inflationary behaviours able to overcome the
shortcomings of Standard Cosmological Model (based on GR). The
related cosmological models seem very realistic and, several
times, capable of matching with the observations
\cite{starobinsky,la}.
  Furthermore, it is possible to show
that, via conformal transformations, the higher--order and
non-minimally coupled terms always correspond to  Einstein gravity
plus one or more than one minimally coupled scalar fields
\cite{teyssandier,maeda,wands,gottloeber}. More precisely,
higher--order terms always
 appear as a contribution of order two in the equations of motion.
 For example, a term like $R^{2}$ gives fourth
 order equations \cite{ruzmaikin}, $R \ \Box R$ gives sixth order
 equations \cite{gottloeber,sixth},
 $R \,\Box^{2}R$ gives eighth order equations \cite{eight} and so on.
  By a conformal transformation,
 any 2nd--order of derivation corresponds to a scalar field: for example,
 fourth--order gravity gives Einstein
 plus one scalar field, sixth order gravity gives Einstein plus two scalar
 fields and so on \cite{gottloeber,schmidt1}.
This feature results very interesting if we want to obtain
multiple inflationary events since an early stage could select
``very'' large-scale structures (clusters of galaxies today),
while a late stage could select ``small'' large-scale structures
(galaxies today) \cite{sixth}. The philosophy is that each
inflationary era is connected with the dynamics of a scalar field.
Furthermore, these extended schemes naturally could solve the
problem of ``graceful exit" bypassing the shortcomings of former
inflationary models \cite{la,aclo}.

However, in the weak-field-limit approximation,  these  theories
are expected to reproduce GR which, in any case, is experimentally
tested only in this limit \cite{will}. This fact is matter of
debate since several relativistic theories
 {\it do not} reproduce exactly Einstein results in the Newtonian
approximation but, in some sense, generalize them. As it was
firstly noticed by Stelle \cite{stelle}, a $R^2$--theory gives
rise to Yukawa--like corrections to the Newtonian potential which
could have interesting physical consequences. For example, some
authors claim to explain the flat rotation curves of galaxies by
using such terms \cite{sanders}. Others \cite{mannheim} have shown
that a conformal theory of gravity is nothing else but a
fourth--order theory containing such terms in the Newtonian limit.
Besides, indications of an apparent, anomalous, long--range
acceleration revealed from the data analysis of Pioneer 10/11,
Galileo, and Ulysses spacecrafts could be framed in a general
theoretical scheme by taking   corrections to the Newtonian
potential into account \cite{anderson}. In general, any
relativistic theory of gravitation can yield corrections to the
Newton potential (see for example \cite{schmidt}) which, in the
post-Newtonian (PPN) formalism, could furnish tests for the same
theory \cite{will}. Furthermore the newborn {\it gravitational
lensing astronomy} \cite{ehlers} is giving rise to additional
tests of gravity over small, large, and very large scales which
very soon will provide direct measurements for the variation of
Newton coupling $G_{N}$ \cite{krauss}, the potential of galaxies,
clusters of galaxies \cite{nottale} and several other features of
gravitating systems. Such data will be, very likely, capable of
confirming or ruling out the physical consistency of GR or of any
ETG.

 In summary, the general feature of ETGs is that the Einstein field
 equations result to be modified
in two senses: $i)$ geometry can be non-minimally coupled to some
scalar field, and/or $ii)$ higher than second order derivative
terms in the metric come out. In the first case, we generically
deal with scalar-tensor theories of gravity or non-minimally
coupled theories; in the second one we deal with higher-order
theories. However several combinations of both situations can
result physically interesting \cite{wands}. In this case, we deal
with higher-order-scalar-tensor theories of gravity.

In all these approaches, the problem of reducing  more general
theories to Einstein standard form has been extensively treated;
one can see that, through a ``Legendre'' transformation on the
metric, higher-order theories, under suitable regularity
conditions on the Lagrangian, take the form of the Einstein one in
which a scalar field (or more than one) is the source of the
gravitational field (see for example \cite{ordsup}
\cite{francaviglia,sokolowski,magnano-soko}); on the other side,
it has been studied the equivalence between models with variable
gravitational coupling with the Einstein standard gravity through
a suitable conformal transformation (see \cite{dicke,nmc}). In any
case, the debate on the physical meaning of conformal
transformations is far to be solved [see \cite{faraoni} and
refererences therein for a comprehensive review]. Several authors
claim for a true physical difference between Jordan frame
(higher-order theories and/or variable gravitational coupling)
since there are experimental and observational evidences which
point out that the Jordan frame could be suitable to better match
 solutions with data. Others state that the true physical frame
is the Einstein one according to the energy theorems
\cite{magnano-soko}. In any case, the discussion is open and no
definite statement has been done up to now. The problem should be
faced from a more general viewpoint and the Palatini approach to
gravity could be useful to this goal.
 The Palatini approach in gravitational theories was
firstly introduced and analyzed by Einstein himself
\cite{palaeinstein}. It was however called Palatini approach as a
consequence of an historical misunderstanding
\cite{buchdahl,frafe}. The fundamental idea at the bases of the
Palatini formalism is to consider the (usually torsion-less)
connection $\Gamma$, entering the definition of the Ricci tensor
to  be independent of the metric $g$ defined on the spacetime $M$.
The Palatini formalisms for the standard Hilbert-Einstein
torsion-less theory results to be equivalent to the purely metric
theory: this follows from the fact that the field equation for the
connection fields states exactly that the same connection
$\Gamma$, firstly considered to be independent, should be the
Levi-Civita connection of the metric $g$. There is consequently no
reason to impose the Palatini variational principle in the
standard Hilbert-Einstein theory instead of the metric (Einstein)
variational principle. The situation however completely changes
when we consider the case of ETGs, depending on analytical
functions of  curvature invariants as $f(R)$, or non-minimally
coupled a scalar field. In these cases, as we will show later in
detail, the Palatini and the metric variational principle provide
different field equations and the theories thus derived surely
differ; see for a partial discussion \cite{FFV} and
\cite{magnano-soko}. The importance of the Palatini approach in
this framework has been recently proven in relation
with  cosmological applications \cite{ABFR,palatinifR}.\\
From a physical viewpoint, considering the metric $g$ and the
connection $\Gamma$ as independent fields is somehow equivalent to
decouple the metric structure of spacetime and its geodesic
structure (i.e. the connection is not the Levi-Civita connection
of $g$), governing respectively the chronological structure of
spacetime and the trajectories of particles, moving in it. This
decoupling enriches the geometrical structure of spacetime and
generalizes the purely metric formalism. This metric-affine
structure of spacetime (here, we  simply mean that a connection
$\Gamma$ and a metric $g$ are involved) is naturally translated,
by means of the same (Palatini) field equations, into a bi-metric
structure of spacetime. Besides the \textit{physical} metric $g$,
another metric $h$ is involved. This new metric, at least in
$f(R)$ theories, is simply related to the connection. As a matter
of facts, the connection $\Gamma$  results to be the Levi-Civita
connection of $h$ and thus provides the geodesic structure of
spacetime. When we consider the case of  non-minimally coupled
interaction in the gravitational Lagrangian (scalar-tensor
theories), the new metric $h$ is somehow related with the
non-minimal coupling. Also in the case of Brans-Dicke like
theories the new metric $h$ can be thus related to a different
geometrical and physical aspect of the gravitational theory.
Thanks to the Palatini formalism this non-minimal coupling effects
and the scalar field, entering the evolution of the gravitational
fields, are separated from the metric structure of spacetime. The
situation mixes when we consider the case of
higher-order-scalar-tensor theories.

In this paper we analyze, through  appropriately defined conformal
transformations, the problem of the equivalence between
higher-order non-minimally coupled theories and  General
Relativity in the  Palatini approach. First, we will do it in the
general context of the field theories and then we reduce to the
cosmological case, that is, we will study the conformal invariance
under the hypotheses of homogeneity and isotropy. In this case, we
also consider the case in which ordinary matter is present,
besides the scalar field, and we make some consideration on the
problem of which is the ``physical system'' between the two
conformally equivalent systems
\cite{brans,francaviglia,sokolowski,magnano-soko}.

The layout of the paper is the following. In Sec.II, we discuss
the conformal transformations through their applications to
non-minimally coupled and higher-order theories of gravity. The
goal is to show that starting from the Jordan frame (at least in
the case in which standard perfect fluid matter is not
considered), through a conformal transformation, the system can
always be reduced to the Einstein frame, where gravity is
minimally coupled to one (or more than one) scalar field(s). In
principle, every ETG is conformally equivalent to GR+scalar
field(s). Sec.III is devoted to the discussion of conformal
transformations in the framework of Palatini approach. Due to the
intrinsic bi-metric structure of such an approach and to the fact
that affine connections coincide with Levi-Civita connection only
in the Hilbert-Einstein case, conformal transformations acquire a
relevant role in order to study chronological and geodesic
structures of spacetime. In other words, they are not only a mere
mathematical tool but they put in evidence that physics could be
different in Einstein and Jordan frame. This last issue is
particularly highlighted in cosmology, as discussed in Sec.IV. In
fact, the solutions derived in the different frames could be
distinguished by observations. This fact shows that Einstein frame
and Jordan frame are physically inequivalent since Palatini field
equations are intrinsically different from those derived in
standard GR. Conclusions are drawn in Sec.V.

\section{Conformal transformations}
Let us start giving detailed examples of conformal transformations
in order to show how they work on the Lagrangian and the field
equations of a given ETG, defined in the Jordan frame. The goal is
to reduce the theory to the Einstein frame, \ie to a minimally
coupled theory plus decoupled scalar field(s). The procedure, in
principle, works for any ETG, but it is extremely useful, as we
will see below, in the interpretation of solutions which, from a
physical viewpoint, should be fitted  against experimental and
observational data. Before discussing conformal transformations in
the framework of the Palatini approach, we work out in detail the
purely metric non-minimally coupled scalar-tensor case and the
$f(R)$ case, giving general considerations at the end of this
section.

\subsection{The Scalar-Tensor case}

In four dimensions, a general  non-minimally coupled scalar-tensor
theory of gravity is given by the effective (purely metric) action
\beq \label{s1} {\cal A}= \int \vol \left[F(\f) R+ \half \gu
\f\ddemu \f\ddenu- V(\f) \right] \eeq where $R$ is the Ricci
scalar, $V(\f)$ and $F(\f)$ are generic functions describing
respectively the potential and the coupling of $\f$.  We shall
adopt Planck units. The Brans-Dicke theory of gravity is a
particular case of the action (\ref{s1}) for $V(\f)$=0
\cite{noibrans}. The variation with respect to $\gd$ gives the
field equations \beq \label{s2} F(\f) G\dmunu= F(\f)\left[R\dmunu-
\half R \gd \right]= -\half T\dmunu- g\dmunu \Box_{g} F(\f)+
F(\f)\ddemunu \eeq which are the generalized Einstein equations;
here $\Box_{\GAM}$ is the d'Alembert operator with respect to the
metric $g$, and $G\dmunu$ is the Einstein tensor. Here and below,
semicolon denotes metric covariant derivatives with respect to
$g$. The energy--momentum tensor relative to the scalar field is
\beq \label{s4} T\dmunu= \f\ddemu \f\ddenu- \half g\dmunu \f\ddea
\f\udea+ g\dmunu V(\f) \eeq The variation with respect to $\f$
provides the Klein--Gordon equation \beq \label{s5} \Box_{g} \f- R
\FDEF(\f)+ \VDEF(\f)= 0 \eeq where $\FDEF= dF(\f)/d\f$, $\VDEF=
dV(\f)/d\f$. This last equation is equivalent to the Bianchi
contracted identity \cite{cqg}. The conformal transformation on
the metric $\gd$ is \beq \label{s6} \gbrd= e^{2 \ome} \gd \eeq in
which $e^{2 \ome}$ is the conformal factor. Under this
transformation,  the Lagrangian density in (\ref{s1}) becomes \beq
\label{s7}
\begin{array}{ll}
\disp{ \sqrt{-g} \left(F R+ \half \gu \f\ddemu \f\ddenu- V\right)}
& = \sqrt{-\gbr} e^{-2 \ome} \left(F \bar{R}- 6 F \Box_{\bar{g}}
\ome+\right. \\
~ & ~ \\
~ & \disp{ \left. -6 F \ome\ddea \ome\udea+ \half \gbru \f\ddemu
\f\ddenu- e^{-2 \ome} V\right)}
\end{array}
\eeq in which $\bar{R}$ and $\Box_{\bar{g}}$ are respectively the
Ricci scalar  and the d'Alembert operator relative to the metric
$\bar{g}$. Requiring the theory in the metric $\gbrd$ to appear as
a standard Einstein theory,  the conformal factor has to be
related to $F$ \cite{nmc}, that is \beq \label{s8} e^{2 \ome}= -2
F. \eeq $F$ must be negative to restore physical coupling. Using
this relation, the Lagrangian density (\ref{s7}) becomes \beq
\label{s9} \sqrt{-g} \left( F R+ \half \gu \f\ddemu \f\ddenu-
V\right) = \sqrt{-\gbr} \left( -\half \bar{R}+ 3 \Box_{\bar{g}}
\ome+ \frac{3 \FDEF^2- F}{4 F^2} \f\ddea \f\udea- \frac{V}{4
F^2}\right), \eeq Introducing a new scalar field $\fbr$ and the
potential $\VBR$, respectively, defined by \beq \label{s10}
\fbr\ddea= \sqrt{\frac{3\FDEF^2- F}{2 F^2}}\, \f\ddea,~~~
\VBR(\fbr(\f))= \frac{V(\f)}{4 F^2(\f)}, \eeq we get \beq
\label{s11} \sqrt{-g} \left( F R+ \half \gu \f\ddemu \f\ddenu-
V\right) = \sqrt{-\gbr} \left( -\half \bar{R}+ \half \fbr\ddea
\fbr\udea- \VBR\right), \eeq which is the usual Hilbert-Einstein
Lagrangian density plus the standard Lagrangian density relative
to the scalar field $\fbr$. (We have not considered the
divergence--type term appearing in the Lagrangian (\ref{s11}); we
will return on this point in our forthcoming considerations).
Therefore, every non-minimally  coupled scalar-tensor theory, in
absence of ordinary matter, \ie perfect fluid,  is conformally
equivalent to an Einstein theory, being the conformal
transformation and the potential suitably defined by (\ref{s8})
and (\ref{s10}). The converse is also true: for a given $F(\f)$,
such that $3 \FDEF^2- F> 0$, we can transform a standard Einstein
theory into a non-minimally coupled scalar-tensor theory. This
means that, in principle, if we are able to solve the field
equations in the framework of the Einstein theory in presence of a
scalar field with a given potential, we should be able to get the
solutions for the scalar-tensor theories, assigned by the coupling
$F(\f)$, via the conformal transformation  (\ref{s8}) with the
constraints given by Eqs.(\ref{s10}). This is exactly what we are
going to discuss in the cosmological context in cases in which the
potentials as well as the couplings are relevant from the point of
view of the fundamental physics. In our opinion, this is not only
a mathematical procedure but, by the Palatini approach, it is
related to the physical degrees of freedom  of the theory which
are, in some sense, ``disentangled"  by the conformal
transformations.

Following the standard terminology, the ``Einstein frame'' is the
framework of the Einstein theory with the minimal coupling and the
 ``Jordan frame'' is the framework of the non-minimally coupled theory
\cite{jordan-fierz}.

We have to make some interesting remarks  with respect to
(\ref{s9}) and (\ref{s10}):  the ``new'' scalar field, defined in
(\ref{s10}) is given in differential form in terms of the ``old''
one and its integration can be not trivial; the second remark
concerns the divergence appearing in (\ref{s9}). The transformed
Lagrangian density obtained from (\ref{s7}) by imposing (\ref{s8})
contains a divergence term, in which  not only the metric but also
its derivative appear, through the connection $\bar{\Gamma}$.
Therefore the equivalence of this total Lagrangian density to the
Hilbert-Einstein Lagrangian density plus scalar field is not
trivial and, due to this fact, the Palatini approach, which
distinguishes a priori the fields $g$ and $\Gamma$, is extremely
relevant. To check that they are actually equivalent, let us
perform the conformal transformation (\ref{s6}) on the field Eqs.
(\ref{s2}), obtaining \beq \label{s15}
\begin{array}{ll}
{\bar{G}}\dmunu & = \disp{ \left( -\frac{1}{2 F}+ \frac{F_{\f
\f}}{F}+ \frac{2 \ome_{\f} \FDEF}{F}- 2 {\ome_{\f}}^2- 2 \ome_{\f
\f} \right)
\f\ddemu \f\ddenu+ } \\
~ & ~ \\
~ & \disp{ +\left( \frac{1}{4 F}- \frac{F_{\f \f}}{F}+
\frac{\ome_{\f} \FDEF}{F}- {\ome_{\f}}^2+ 2 \ome_{\f \f} \right)
\gbrd \f\ddea \f\udea+ \left( -\frac{F_{\f}}{F}+ 2
\ome_{\f}\right) \gbrd \Box_{\bar{\Gamma}}
\f+} \\
~ & ~ \\
~ & \disp{ +\left( \frac{F_{\f}}{F}- 2 \ome_{\f}\right)
(\nabla^{\bar{\Gamma}})_\mu (\nabla^{\bar{\Gamma}})_\nu \f-
\frac{1}{2 F} e^{-2 \ome} \gbrd V},
\end{array}
\eeq in which $(\nabla^{\bar{\Gamma}})_\mu$ is the covariant
derivative with respect to $x\umu$ relative to the connection
$\bar{\Gamma}$ and $\Box_{\bar{\Gamma}}$ is nothing but
$\Box_{\bar{g}}$. If $\ome$ satisfies the relation \beq
\label{s16} \frac{\FDEF}{F}- 2 \ome_\f= 0, \eeq Eqs. (\ref{s15})
can be rewritten as \beq \label{s17} {\bar{G}}\dmunu= \frac{3
\FDEF^2- F}{2 F^2} \f\ddemu \f\ddenu- \gbrd \frac{3 \FDEF^2- F}{2
F^2} \f\ddea \f\udea- \gbrd \frac{e^{-2 \ome}}{2 F} V. \eeq Then,
using the transformations (\ref{s10}) and defining the potential
\beq \label{s18} W(\fbr(\f))= -\frac{e^{-2 \ome(F)}}{2 F} V, \eeq
where $\ome(F)$ satisfies (\ref{s16}), Eq. (\ref{s17}) becomes
\beq \label{s19} {\bar{G}}\dmunu= \fbr\ddemu \fbr\ddenu- \half
\gbrd \fbr\ddea \fbr\udea- \gbrd W, \eeq which correspond to the
Einstein field equations in presence of a scalar field $\fbr$ with
potential $W$. The function $\ome(F)$ is obtained from
(\ref{s16}), that is \beq \label{s20} \ome= \half \ln{F}+ \ome_0
\eeq in which $\ome_0$ is an integration constant. The potential
$W$ takes the form \beq \label{s21} W= -\frac{V}{2 \xi F}. \eeq
From (\ref{s21}) and the second of (\ref{s10}), we see that,
fixing $\xi= -2$, the definition of $W$ coincides with that one of
$\VBR$. We have then the full compatibility with the Lagrangian
approach obtaining for $\ome$ the  relation (\ref{s8}); in this
sense, the equivalence between the non-minimally coupled
Lagrangian density and the Hilbert-Einstein Lagrangian density
plus scalar field is verified.

A final remark regards Eqs.(\ref{s10}): actually, from (\ref{s9})
the relation between $\fbr\ddea$ and $\f\ddea$ present a $\pm$
sign in front of the square root, which corresponds to have the
same or opposite sign in the derivative of $\f$ and $\fbr$ with
respect to $x\da$. This ambiguity acquire a physical meaning in
the interpretation of the scalar field, as we shall see below.

\subsection{Higher-order gravity case }

In general, fourth-order theories of gravity are given by the
action \beq\label{h1} {\cal A}=\int\vol f(R)\,, \eeq where $f(R)$
is an analytic function of the Ricci curvature scalar $R$. We are
considering the simplest case of fourth-order gravity but we can
construct such kind of theories also using the invariants
$R\dmunu$ o $R\ua_{\gamma\mu\nu}$. However, for cosmological
considerations, theories like (\ref{h1}) are sufficiently general
\cite{mijic}. Hilbert--Einstein action is recovered for $f(R)=R$.
Varying with respect to $g\dab$, we get the field equations
\beq\label{h2}
f'(R)R\dab-\frac{1}{2}f(R)g\dab=f'(R)^{;\umunu}\left(
g_{\alpha\mu}g_{\beta\nu}-g\dab\gd\right)\,, \eeq which are
fourth-order equations thanks to the term $f'(R)^{;\mu\nu}$. The
prime  indicates the derivative with respect to  $R$. Putting in
evidence the Einstein tensor, we have \beq\label{h3}
G\dab=\frac{1}{f'(R)}\left\{\frac{1}{2}g\dab\left[f(R)-Rf'(R)\right]
+f'(R)\ddeab -g\dab\Box f'(R)\right\}\,, \eeq where the
gravitational contributions in the stress-energy tensor can be
interpreted, via conformal transformations,  as  scalar field
contributions and then as ``matter" terms. Performing the conformal
transformation (\ref{s6}) , we get \beq\label{h4}
\bar{G}\dab=\frac{1}{f'(R)}\left\{\frac{1}{2}g\dab\left[f(R)-Rf'(R)\right]
+f'(R)\ddeab- g\dab\Box f'(R)\right\}+ \eeq
$$ +2\left(\omega_{;\alpha ;\beta}
+g\dab\Box \omega -\omega\ddeab+
\frac{1}{2}g\dab\omega_{;\gamma}\omega^{;\gamma}\right)\,.$$ We
can choose the conformal factor \beq\label{h5}
\omega=\frac{1}{2}\ln |f'(R)|\,, \eeq which has to be substituted
into (\ref{h4}). Rescaling $\omega$ in such a way that
\beq\label{h6} k\phi =\omega\,, \eeq and $k=\sqrt{1/6}$, we obtain
the Lagrangian equivalence \beq \label{h7} \sqrt{-g} f(R)=
\sqrt{-\gbr} \left( -\half \bar{R}+ \half \fbr\ddea \fbr\udea-
\VBR\right) \eeq and the Einstein equations in standard form
\beq\label{h8} \bar{G}\dab=
\phi\ddea\phi\ddeb-\frac{1}{2}\bar{g}\dab\phi_{;\gamma}\phi^{;\gamma}
+\bar{g}\dab V(\phi)\,, \eeq  with the potential \beq\label{h9}
V(\phi)=\frac{e^{-4k\phi}}{2}\left[f(\phi)-{\cal
F}\left(e^{2k\phi}\right)e^{2k\phi}\right]
=\frac{1}{2}\frac{f(R)-Rf'(R)}{f'(R)^{2}}\,. \eeq
 ${\cal F}$ is the inverse function of
 $f'(\phi)$ and $f(\phi)=\int \exp (2k\phi) d{\cal F}$. However, the problem is
 completely solved if
$f'(\phi)$ can be analytically inverted. In summary, a
fourth-order theory is conformally equivalent to the standard
second-order Einstein theory plus a scalar field (see also
\cite{francaviglia,ordsup}).

If the theory is higher than fourth order, we have Lagrangian
densities of the form \cite{buchdahl,gottloeber,sixth},
\beq\label{h10} {\cal L}={\cal L}(R,\Box R,...\Box^{k} R)\,. \eeq
Every $\Box$ operator introduces two further terms of derivation
into the field equations. For example a theory like
\beq\label{h11} {\cal L}=R\Box R\,, \eeq is a sixth-order theory,
and the above approach can be pursued considering a conformal
factor of the form
 \beq\label{h12} \omega=\frac{1}{2}\ln \left|\frac{\pa
{\cal L}}{\pa R} +\Box\frac{\pa {\cal L}}{\pa \Box R}\right|\,.
\eeq In general,  increasing two orders of derivation in the field
equations (\ie every term $\Box R$), corresponds to add a scalar
field in the conformally transformed frame \cite{gottloeber}. A
sixth-order theory can be reduced to an Einstein theory with two
minimally coupled scalar fields; a $2n$-order theory can be, in
principle, reduced to an Einstein theory + $(n-1)$-scalar fields.
On the other hand, these considerations can be directly
generalized to higher-order-scalar-tensor theories in any number
of dimensions as shown in \cite{maeda}.

As concluding remarks, we can say that conformal transformations
works at three levels: $i)$ on the Lagrangian of the given
ETG-theory; $ii)$ on the field equations; $iii)$ on the solutions.
The table below summarizes the situation for fourth-order gravity
(FOG), non-minimally coupled scalar-tensor theories (NMC) and
standard Hilbert-Einstein (HE) theory. Clearly, direct and inverse
transformations correlate all the steps of the table but no
absolute criterion, at this point of the discussion, is capable of
stating what is the ``physical" framework since from a mathematical
point of view all the frames are equivalent (see also
\cite{magnano-soko} for a detailed discussion). The Palatini
approach can aid in this task.

\begin{center}
\begin{tabular}{|ccccc|} \hline
  ${\cal L}_{FOG}$ & $\longleftrightarrow$ & ${\cal L}_{NMC}$ & $\longleftrightarrow$ &
     ${\cal L}_{HE}$ \\
  $\updownarrow$ &  & $\updownarrow$ &  & $\updownarrow$ \\
  FOG Eqs. & $\longleftrightarrow$ & NMC Eqs. & $\longleftrightarrow$ & Einstein Eqs. \\
  $\updownarrow$ &  & $\updownarrow$ &  & $\updownarrow$ \\
  FOG Solutions & $\longleftrightarrow$ & NMC Solutions & $\longleftrightarrow$ &
  Einstein Solutions \\ \hline
\end{tabular}
\end{center}

\section{Palatini Approach and Conformal Transformations}
As we said, the Palatini approach, considering $g$ and $\Gamma$ as
independent fields, is ``intrinsically" bi-metric and capable of
disentangle the geodesic from the chronological structure of a
given manifold. Starting from these features for ETG, conformal
transformations assume a fundamental role in defining the affine
connection which is merely ``Levi-Civita" only for Hilbert-Einstein
gravity. In this section, we work out examples showing how
conformal transformations assume a fundamental physical role.

\subsection{ $f(R)$ gravity in Palatini approach and the intrinsic conformal structure}
Let us start from the case of fourth-order gravity where Palatini
variational principle is straightforward in showing the
differences with Hilbert-Einstein variational principle, involving
only metric. Besides cosmological applications of $f(R)$ gravity
have shown the importance of the Palatini formalism in this
framework, giving physically relevant results and avoiding
singular behaviors of solutions \cite{ABFR,palatinifR}. This last
nice feature is not present in the standard metric approach.  The
standard and more general $f(R)$ Lagrangian suitable for our
considerations is
\begin{equation} \label{lagrfR}
A=A_{\mathrm{grav}}+A_{\mathrm{mat}}=\int \sqrt{-g} \; \; [
f(R)+2\kappa L_{\mathrm{mat}}(\Psi) ]  d^{4}x
\end{equation}
where $R\equiv R( g,\Gamma) =g^{\alpha\beta}R_{\alpha \beta}(\Gamma )$ is
the \textit{generalized Ricci scalar} and $
R_{\mu \nu }(\Gamma )$ is the Ricci tensor of a torsionless
connection $\Gamma$, which a priori has no relations with the metric $g$ of spacetime.
The gravitational part of the Lagrangian is controlled by a given real analytical
function of one real variable $f(R)$,
 while $\sqrt{-g}$ denotes a related scalar density of weight $1$.
The  Lagrangian contains also a matter part, usually chosen to be
the \textit{Lagrangian} of the perfect fluid $L_{\mathrm{mat}}$ in
minimal interaction with the gravitational field but it can be
also a minimally coupled scalar field(s) Lagrangian. This
Lagrangian is dependent on
 matter fields $\Psi$ together with their first
 derivatives and  equipped with a gravitational coupling constant
 $\kappa=8\pi G$ which we restore now, with respect to the previous
 considerations for the sake of clearness in the discussion.
More general couplings between the gravitational Lagrangian and
matter fields, involving the covariant derivatives with respect to
$\Gamma$ of matter fields could be considered. We remark, however,
that the absence of these interactions do not change much the
physics of the theory, owing to the conformal relation in the
bi-metric structure of spacetime in $f(R)$ gravity; see
\cite{ABFR} for details.   Field equations, deriving from the
Palatini variational principle are (we assume the  spacetime
manifold to be a Lorentzian manifold ${\cal M}$ with dim${\cal
M}=4$; see \cite{FFV}):
\begin{equation}
f^{\prime }(R)R_{(\mu\nu)}(\Gamma)-\frac{1}{2}f(R)g_{\mu \nu
}=\kappa T_{\mu\nu}\label{ffv1}
\end{equation}
\begin{equation}
\nabla _{\alpha }^{\Gamma }(\sqrt{-g}f^{\prime}(R)g^{\mu \nu })=0
\label{ffv2}
\end{equation}
where ${\displaystyle T_{\mu\nu}=-{2}  \frac{\delta
L_{\mathrm{mat}}}{\delta g_{\mu\nu}}}$ denotes the matter source
stress-energy tensor and $\nabla^{\Gamma}$ is the covariant
derivative with respect to $\Gamma$.  We shall use the standard
notation denoting by $R_{(\mu\nu)}$ the symmetric part of
$R_{\mu\nu}$, \ie $R_{(\mu\nu)}\equiv
\frac{1}{2}(R_{\mu\nu}+R_{\nu\mu)})$. In order to get (\ref{ffv2})
one has to additionally assume that $L_{\mathrm{mat}}$ is
functionally independent of $\Gamma$ (as already remarked);
however it may contain metric  covariant derivatives
$\stackrel{g}{\nabla}$ of fields. This means that the matter
stress-energy tensor $T_{\mu\nu}=T_{\mu\nu}(g,\Psi)$ depends on
the metric $g$ and some matter fields denoted here by $\Psi$,
together with their derivatives (covariant derivatives with
respect to the Levi-Civita connection of $g$). From (\ref{ffv2})
one sees that $\sqrt{-g}f^{\prime }(R)g^{\mu \nu }$ is a symmetric
twice contravariant tensor density of weight $1$. As previously
discussed in \cite{ABFR} and \cite{FFV} this naturally lead us to
define a new metric $h_{\mu \nu}$, such that the following
relation holds true:
\begin{equation}\label{h_met}
\sqrt{-g}f^{\prime }(R)g^{\mu \nu}=\sqrt{-h}h^{\mu \nu }
\end{equation}
This \textit{ansatz} is suitably made in order to impose $\Gamma$
to be the Levi-Civita connection of $h$ and the only restriction
is that $\sqrt{-g}f^{\prime }(R)g^{\mu \nu}$ should be
non-degenerate. In the case of Hilbert-Einstein Lagrangian,
$f^{\prime}(R)=1$ and the statement is trivial. The above equation
(\ref{h_met}) imposes that the two metrics $h$ and $g$ are
conformally equivalent. The corresponding conformal factor can be
easily found to be $f^{\prime}(R)$ (in dim${\cal M}=4)$ and the
conformal transformation results to be ruled by:
\begin{equation}\label{h_met1}
h_{\mu \nu }=f^{\prime}(R)g_{\mu \nu }
\end{equation}
Therefore, as it is well known, equation (\ref{ffv2}) implies that
$\Gamma =\Gamma _{LC}(h)$ and $R_{(\mu\nu)}(\Gamma)=R_{\mu \nu }(h)\equiv R_{\mu\nu}$.
Field equations can be supplemented by the scalar-valued equation
obtained by taking the trace of (\ref{ffv1}),
(we define $\tau=\mathrm{tr}\hat T$)
\begin{equation} \label{structR}
f^{\prime }(R)R-2f(R)=\kappa g^{\alpha\beta}T_{\alpha\beta}\equiv \kappa\tau
\end{equation}
which controls solutions of (\ref{ffv2}). We shall refer to this
scalar-valued equation as the \textit{structural equation} of
spacetime. In the vacuum case (and radiating spacetimes, such that
$\tau=0$) this scalar-valued equation admits only constant
solutions and the universality of Einstein field equations holds
\cite{FFV}, corresponding to a theory with constant cosmological
constant \cite{cosmconst}. In the case of interaction with matter
fields, the structural equation (\ref{h_met1}), if explicitly
solvable, provides an expression of $R=F(\tau)$ and consequently
both $f(R)$ and $f^\prime (R)$ can be expressed in terms of
$\tau$. The matter content of spacetime thus rules the bi-metric
structure of spacetime and consequently both the geodesic and
metric structures \cite{ABFR} which are intrinsically different.
This behavior generalizes the vacuum case and  corresponds to the
case of a time varying cosmological constant. In other words, due
to these features, conformal transformations, which allow to pass
from a metric structure to another one, acquire an intrinsic
physical meaning since ``select" metric and geodesic structures
which for a given ETG, in principle, {\it do not} coincide.

\subsection{The case of Scalar-Tensor Gravity}
Let us now try to extend the above formalism to  case of
non-minimally coupled scalar-tensor theories. The effort is to
understand if and how the bi-metric structure of spacetime behaves
in this cases and which could be its geometrical and physical
interpretation. As a general result, the Palatini formalism and
the bi-metric structure ``select" intrinsically different theories
of gravity. The importance of these theories
 is well known in cosmological applications.
For completeness reasons we start by considering scalar-tensor
theories in the Palatini formalism, calling $A_1$ the action
functional. After, we take into account the case of decoupled
non-minimal interaction between a scalar-tensor theory and a
$f(R)$ theory, calling $A_2$ this action functional. We finally
consider the case of non-minimal-coupled interaction between the
scalar field $\phi$ and the gravitational fields $(g, \Gamma)$,
calling $A_3$ the corresponding action functional. Particularly
significant is, in this case, the limit  of low curvature $R$.
This resembles the physical relevant case of present values of
curvatures of the universe and it is important for cosmological
applications.

The  action (\ref{s1}) for scalar-tensor gravity  can be
generalized, in order to better develop the Palatini approach, as:
\begin{equation} \label{lagrfR1}
A_1=\int \sqrt{-g} \; \; [ F(\phi) R+{\epsilon \over 2}
\stackrel{g}{\nabla}_\mu \phi  \stackrel{g}{\nabla}^{ \mu} \phi
-V(\phi)+ \kappa L_{\mathrm{mat}}(\Psi, \stackrel{g}{\nabla} \Psi)
]  d^{4}x
\end{equation}
where $\phi$, as above, is an arbitrary scalar field. The values
of $\epsilon=\pm 1$ selects between standard scalar field theories
and quintessence field theories. The relative ``signature" can be
selected by conformal transformations. Field equations for the
gravitational part of the action are, respectively for the metric
$g$ and the connection $\Gamma$:
\begin{equation}\label{40}
\cases{ F(\phi) [R_{(\mu\nu)}-{1 \over 2} R g_{\mu \nu} ] = \kappa
[T^\phi_{\mu \nu }+T^\mathrm{mat}_{\mu \nu }]  \cr \nabla _{\alpha
}^{\Gamma }(\sqrt{-g} F (\phi)g^{\mu \nu })=0 }
\end{equation}
where we have defined the stress-energy tensors for the scalar
field and the matter Lagrangian, respectively as:
\begin{equation} \label{tmunudef}
\cases{ T^\phi_{\mu \nu }=-\frac{2}{\kappa }  \frac{\delta
L_{\phi}}{\delta g_{\mu\nu}} \cr T^\mathrm{mat}_{\mu \nu}=-{2}
\frac{\delta L_{\mathrm{mat}}}{\delta g_{\mu\nu}} }
\end{equation}
and $R_{(\mu\nu)}$ is the same defined in (\ref{ffv1}).
For matter fields we have the following field equations:
 \begin{equation}\label{42}
\cases{ \epsilon \Box \phi= - V_{\f} (\phi) + F_{\f} (\phi) R \cr
\frac{\delta L_{\mathrm{mat}}}{\delta \Psi}=0 }
\end{equation}
In this case, the structural equation of spacetime implies that:
 \begin{equation} \label{stru1a}
R=-\frac{\kappa (\tau^\phi+\tau^\mathrm{mat})}{F(\phi)}
\end{equation}
which expresses the value of the Ricci scalar curvature in terms
of traces of the stress-energy tensors of matter (we have to
require $F(\phi) \ne 0$). The bi-metric structure of spacetime is
thus defined by the ansatz:
 \begin{equation}\label{41}
\sqrt{- g} F (\phi) g^{\mu \nu }=\sqrt{- h} h^{\mu \nu }
 \end{equation}
such that $g$ and $h$ result to be conformal:
 \begin{equation} \label{bimetri1}
h_{\mu \nu }=  F (\phi) g_{\mu \nu }
 \end{equation}
The conformal factor is exactly the  minimal interaction factor.
We remark that from (\ref{stru1a}) it follows that in the vacuum
case $\tau^\phi=0$ and $\tau^\mathrm{mat}=0$ this theory is
equivalent to the standard Einstein one without matter. On the
other hand, for $F(\phi)=F_0$ we recover the Einstein theory plus
a minimally coupled scalar field. As last remark and keeping in
mind the discussion of the previous section, the Palatini approach
intrinsically furnishes the conformal structure (\ref{bimetri1})
of the theory which is trivial in the Einstein, minimally coupled
case.

\subsection{Decoupled non-minimal interaction in scalar-tensor $f(R)$ theories}

As a further step, we generalize the results of previous
subsection, considering the case of a non-minimal coupling in the
framework of $f(R)$ theories. The action functional can be written
as:
\begin{equation} \label{lagrfR2}
A_2=\int \sqrt{ -g} \; \; [ F(\phi) f(R)+{\epsilon \over 2}
\stackrel{g}{\nabla}_\mu \phi  \stackrel{g}{\nabla}^{ \mu} \phi
-V(\phi)+ \kappa L_{\mathrm{mat}}(\Psi, \stackrel{g}{\nabla} \Psi)
]  d^{4}x
\end{equation}
where $f(R)$ is, as usual, any analytical function of the  Ricci
scalar $R$. Field equations (in the Palatini formalism) for the
gravitational part of the action are:
\begin{equation}
\cases{ F(\phi) [f^\prime (R ) R_{(\mu\nu)}-{1 \over 2} f(R)
g_{\mu \nu} ] = \kappa [T^\phi_{\mu \nu }+T^\mathrm{mat}_{\mu \nu
}]  \cr \nabla _{\alpha }^{\Gamma }(\sqrt{ -g} F(\phi)
f^{\prime}(R) g^{\mu \nu })=0 }
\end{equation}
where we have defined the stress-energy tensors for the scalar
field and the matter Lagrangian, respectively as in
Eqs.(\ref{tmunudef}). For scalar and matter fields we have
otherwise the following field equations:
 \begin{equation}
\cases{ \epsilon \Box \phi= - V_{\f} (\phi) + \sqrt{ -g} F_{\f}
(\phi) f(R) \cr \frac{\delta L_{\mathrm{mat}}}{\delta \Psi}=0 }
\end{equation}
where the non-minimal interaction term enters into the modified
Klein-Gordon equations. In this case the structural equation of
spacetime implies that:
 \begin{equation} \label{stru1}
f^\prime (R) R-2 f(R)=\frac{\kappa
(\tau^\phi+\tau^\mathrm{mat})}{F(\phi)}
\end{equation}
We remark that this equation, if solved,  expresses the value of
the Ricci scalar curvature in terms of traces of the stress-energy
tensors of matter (we have to require again that $F(\phi) \ne 0$).
The bi-metric structure of spacetime is thus defined by the
ansatz:
 \begin{equation}\label{47}
\sqrt{- g} F (\phi) f^\prime (R) g^{\mu \nu }=\sqrt{- h} h^{\mu
\nu }
 \end{equation}
such that $g$ and $h$ result to be conformally related by:
 \begin{equation}
h_{\mu \nu }=  F (\phi) f^\prime (R) g_{\mu \nu }
 \end{equation}
We remark that, once the structural equation is solved, the
conformal factor depends  on the values of the matter fields
 ($\phi, \Psi$) or, more precisely, on the traces of the
  stress-energy tensors and the value of $\phi$.
From equation (\ref{stru1}), it follows that in the case of
vacuum, \ie in the case that both $\tau^\phi=0$ and
$\tau^\mathrm{mat}=0$, the universality of Einstein field equation
still holds as in the case  of minimally interacting $f(R)$
theories \cite{FFV}. The proof is very simply, as it follows
exactly the headlines of \cite{FFV}. The validity of this property
is related to the decoupling of the scalar field and the
gravitational field in this case.

\subsection{The general case}
Let us finally consider the case where the gravitational
Lagrangian is a general function of $\phi$ and $R$. The action
functional can thus be written as:
\begin{equation} \label{lagrfR3}
A_3=\int \sqrt{ -g} \; \; [ K(\phi,R)+{\epsilon \over 2}
\stackrel{g}{\nabla}_\mu \phi  \stackrel{g}{\nabla}^{ \mu} \phi
-V(\phi)+ \kappa L_{\mathrm{mat}}(\Psi, \stackrel{g}{\nabla} \Psi)
]  d^{4}x
\end{equation}
Field equations for the gravitational part of the action are:
\begin{equation}
\cases{ \left[ \frac{\partial \; K(\phi,R)}{\partial R}  \right]
R_{(\mu\nu)}-{1 \over 2} K(\phi, R)  g_{\mu \nu}  = \kappa
[T^\phi_{\mu \nu }+T^\mathrm{mat}_{\mu \nu }]  \cr \nabla _{\alpha
}^{\Gamma } \left( \sqrt{ -g} \left[ \frac{\partial \;
K(\phi,R)}{\partial R} \right] g^{\mu \nu } \right) =0 }
\end{equation}
where we have defined the stress-energy tensors for the scalar
field and the matter Lagrangian, respectively as in
Eqs.(\ref{tmunudef}). For matter fields we have the following
field equations:
 \begin{equation}
\cases{ \epsilon \Box \phi= - V_{\f} (\phi) + \left[
\frac{\partial \; K(\phi,R)}{\partial \phi }  \right] \cr
\frac{\delta L_{\mathrm{mat}}}{\delta \Psi}=0 }
\end{equation}
The structural equation of spacetime can be expressed as:
 \begin{equation} \label{stru2}
\frac{\partial K(\phi, R)}{\partial R} R-2 K(\phi, R) =\kappa
(\tau^\phi+\tau^\mathrm{mat})
\end{equation}
We remark that this equation, if  solved,  expresses again the
form of the Ricci scalar curvature in terms of traces of the
stress-energy tensors of matter (we should impose regularity
conditions and, for example, $K(\phi,R) \ne 0$). The bi-metric
structure of spacetime is thus defined by the ansatz:
 \begin{equation}
\sqrt{- g} \frac{\partial K(\phi, R)}{\partial R}   g^{\mu \nu
}=\sqrt{- h} h^{\mu \nu }
 \end{equation}
such that $g$ and $h$ result to be conformally related by
 \begin{equation} \label{conf3}
h_{\mu \nu }=  \frac{\partial K(\phi, R)}{\partial R}  g_{\mu \nu
}
 \end{equation}
We remark again that, once the structural equation is solved, the
conformal factor depends just on the values of the matter fields
and (the trace of) their stress energy tensors. In other words,
the evolution, the definition of the conformal factor and the
bi-metric structure is ruled by the values of traces of the
stress-energy tensors and by the value of the scalar field $\phi$.
In this case, the universality of Einstein field equations does
not hold anymore in general.
 This is evident from (\ref{stru2}) where the strong coupling between $R$ and $\phi$ avoids
 the possibility, also in the vacuum case, to achieve also constant simple  solutions for the
  structural equations (\ref{stru2}).
We consider furthermore the case when small values of $R$ are
considered, corresponding to the very important cases of small
curvature spacetimes. As already explained, this limit represent,
as a good approximation, the present epoch of the observed
universe under suitably regularity conditions. A Taylor expansion
of the analytical function $K(\phi, R)$ can be performed:
 \begin{equation}
 K(\phi, R)=K_0 (\phi)+K_1 (\phi) R+o(R^2)
 \end{equation}
where only the first leading term in $R$ is considered and we have defined:
 \begin{equation}
\cases{ K_0 (\phi)=  K(\phi, R)_{R=0}\cr K_1 (\phi)=\left(
\frac{\partial K(\phi, R)}{ \partial R } \right)_{R=0}}
 \end{equation}
Substituting this expression in (\ref{stru2}) and (\ref{conf3}) we
get (neglecting higher order approximations in $R$) the structural
equation and the bi-metric structure in this particular case. From
the structural equation we get:
  \begin{equation} \label{Rgenapr}
 R=\frac{1}{K_1 (\phi) } [ -\kappa  (\tau^\phi+\tau^\mathrm{mat})- 2 K_0 (\phi)]
 \end{equation}
such that the value of the Ricci scalar is always determined, in
this first order approximation, in terms of
$(\tau^\phi,\tau^\mathrm{mat},\phi)$. The bi-metric structure is
otherwise simply defined by means of the first term of the Taylor
expansion. We have:
\begin{equation}
h_{\mu \nu }=  K_1 (\phi) g_{\mu \nu }
\end{equation}
which reproduces, as expected, the scalar-tensor case
(\ref{bimetri1}).  In other words, scalar-tensor theories  can be
recovered in a first order approximation of a general theory where
gravity and non-minimal couplings are any (compare (\ref{Rgenapr})
with (\ref{stru1})). This fact agrees with the above
considerations where Lagrangians of physical interactions are
stochastic functions with local gauge invariance properties
\cite{ottewill}.

\section{Cosmological applications}
The above discussion tells us that, for a given ETG, Palatini
approach intrinsically define a bi-metric structure where geodesic
and chronological structures of spacetime do not coincide a
priori. This fact is extremely relevant in the interpretation of
conformal transformations since the interpretation of physical
results in the metrics $h\dmunu$ and $g\dmunu$ (or alternatively
$\bar{g}\dmunu$ and $g\dmunu$)is something different since, in the
Palatini formalism $h$ and $g$ are entangled. This means that $g$
provides the chronological structure while $h$ is related to the
geodesic structure as the affine connection is assumed to be
$\Gamma=\Gamma_{LC}(h)$. This feature assume a crucial role at the
level of the solutions which can be worked out in the two
dynamics, first of all in cosmology. In fact, a bad interpretation
of the geodesic structure of a given spacetime can lead to
misunderstand the results and the interpretation of observations.
In this section, we want to show how the ``same" theory,
conformally transformed, can give rise to completely different
cosmological solutions. For example, in the Einstein frame we can
have solutions with cosmological constant which is the same at
every epoch while in the Jordan frame a self-interacting potential
and a non-minimal coupling come out. This fact leads to a
completely different interpretation of data. The shortcoming is
unambiguously solved only if the structure of affine connections
is completely controlled as in the Palatini approach.

In order to support these statements, let us take into account
scalar-tensor theories in the the Friedmann-Robertson-Walker
cosmology. A part the interest of such theories discussed in the
Introduction, they are remarkable since, as we have seen,
represent the low-curvature limit of general non-minimally coupled
higher-order theories whose interpretation is straightforward in
the Palatini approach.

 Let us assume now that the spacetime
manifold is described by a FRW metric. The Lagrangian density
(\ref{s1}) takes the form  \beq \label{c1} L_t= 6 F(\f) a \ad^2+ 6
\FDEF(\f) a^2 \ad \fd- 6 F(\f)a K + \half a^3 \fd^2- a^3 V(\f).
\eeq  With the subscript $t$, we mean that the time--coordinate
considered is the cosmic time $t$: this remark is important for
the forthcoming discussion. Here $a$ is the scale factor of the
universe and $K$ is spatial curvature constant. The
Euler--Lagrange equations relative to (\ref{c1}) are then \beq
\label{c2} \left \{
\begin{array}{l}
\disp{ \frac{2 \add}{a}+ \frac{\ad^2}{a^2}+ \frac{2 \FDEF \ad
\fd}{F a}+ \frac{\FDEF \fdd}{F}+ \frac{K}{a^2}+ \frac{F_{\f \f}
\fd^2}{F}-
\frac{\fd^2}{4 F}+ \frac{V}{2 F}= 0} \\
~ \\
\disp{ \fdd+ \frac{3 \ad \fd}{a}+ \frac{6 \FDEF \ad^2}{a^2}+
\frac{6 \FDEF \add}{a}+ \frac{6 \FDEF K}{a^2}+ \VDEF= 0}
\end{array}
\right. \eeq which correspond to the (generalized) second order
Einstein equation and to the Klein--Gordon equation in the FRW
case. The energy function relative to (\ref{c1}) is \beq
\label{c3} E_t  =\frac{\pa L_t}{\pa \ad} \ad+ \frac{\pa L_t}{\pa
\fd} \fd- L_t= 6 F a \ad^2+ 6 \FDEF a^2 \ad \fd+ 6 F a K+ \half
a^3 \fd^2+ a^3 V =0\eeq which is the first order generalized
Einstein equation.

 Performing the conformal transformation
defined by (\ref{s6}), (\ref{s8}), (\ref{s10}) on the FRW metric,
one should obtain the corresponding expression for the Lagrangian
and the corresponding equations of the Einstein--cosmology from
the nonstandard coupled Lagrangian (\ref{c1}) and from the
generalized Einstein and Klein--Gordon equations, respectively.
Unfortunately we see that the presence of the conformal factor
(\ref{s8}) implies that the transformed line element which is
obtained is no longer expressed in the ``cosmic time form''.
Actually the scale factor of the Einstein theory can be defined as
the scale factor of the non-minimally coupled theory multiplied by
the conformal factor, but the time coordinate of the Einstein
theory has to be redefined if we require to have the cosmic time
as well. Absorbing the conformal factor in the redefinition of
time, we obtain the transformation on the time coordinate.
Therefore, the transformation from the Jordan frame to the
Einstein frame in the cosmological case is given by \beq
\label{c4} \left \{
\begin{array}{l}
\abr= \sqrt{-2 F(\f)}\, a \\
~ \\
\disp{ \frac{d\fbr}{dt}= \sqrt{\frac{3 \FDEF^2- F}{2 F^2}}\,
\frac{d\f}{dt}} \\
~ \\
d\tbr= \sqrt{-2 F(\f)}\, dt.
\end{array}
\right. \eeq From the Palatini point of view, these
transformations are ``natural" due to the intrinsic different
geodesic structure of the two frames. Furthermore, the system of
Eqs.(\ref{c2}),(\ref{c3}) and the relations (\ref{c4}) to pass
from the Jordan frame to the Einstein frame are immediately
recovered from the Palatini field equations (\ref{40}) and
(\ref{42}), linked together by the structural equation
(\ref{stru1}). Moreover in the Palatini formalism, the
redefinition of cosmic time in the two frames (\ie considering $h$
or $g$ as the physical metric) naturally follows from (\ref{41})
and reproduces (\ref{c4}).In other words, Palatini field equations
give, at once, dynamics of fields and, being endowed with a
bi-metric structure, the relation between the Jordan frame and the
Einstein frame.

Using the first and the third of (\ref{c4}), the scale factor
$\bar{a}$ in the Einstein frame depends only on $\bar{t}$. The
factor $F(\phi)$, which modifies the geodesic structure, is
absorbed into the definition of the cosmic time in the Einstein
frame. The second of (\ref{c4}) corresponds to the first of
relations (\ref{s10}) under the given assumption of homogeneity
and isotropy. Under transformation (\ref{c4}) we have that \beq
\label{c5}
\begin{array}{ll}
\disp{ \frac{1}{\sqrt{-2 F}} L_t} & \disp{ =\frac{1}{\sqrt{-2 F}}}
\left( 6 F a \ad^2+ 6 \FDEF a^2 \ad \fd- 6 F a K+ \half a^3 \fd^2-
a^3 V\right)=
\\
~ & ~ \\
~ & = \disp{ -3 \abr \abrd^2+ 3 K \abr+ \half \abr^3 \fbrd^2-
\abr^3 \VBR(\fbr)= \LBR_\tbr}
\end{array}
\eeq in which the dot over barred quantities means the derivative
with respect to $\tbr$; $L_t$ is given by (\ref{c1}) and
$\LBR_\tbr$ coincides with the ``point--like'' Lagrangian obtained
from the Hilbert-Einstein action plus a scalar field under the
assumption of homogeneity and isotropy. In this way, the
invariance of the homogeneus and isotropic action under (\ref{c4})
is restored, being $L_t$ and $\LBR_\tbr$ equivalent by the
(\ref{c5}). The same correspondence as (\ref{c5}) exists between
the energy function $E_t$ and $\EBR_\tbr$, that is, there is
correspondence between the two first order Einstein equations in
the two frames. It is interesting to note that the relation
(\ref{c5}) reflects the Palatini bi-metric structure: the
Lagrangians are equivalent only if the time is conformally
transformed and Levi-Civita connection is restored in the new
metric.

We focus now our attention on the way in which the Euler--Lagrange
equations transform under (\ref{c4}). The Euler--Lagrange
equations relative to (\ref{c5}) are the usual second order
Einstein equation and Klein--Gordon equation \beq \label{c6} \left
\{
\begin{array}{l}
\disp{ \frac{2 \abrdd}{\abr}+ \frac{\abrd^2}{\abr}+
\frac{K}{\abr^2}+ \half \fbrd^2- \VBR= 0} \\
~ \\
\disp{ \fbrdd+ \frac{3 \abrd \fbrd}{\abr}+ \VBRDEF= 0}.
\end{array}
\right. \eeq Under (\ref{c4}) it is straightforward to verify that
they become \beq \label{c7} \left \{
\begin{array}{l}
\disp{ \frac{2 \add}{a}+ \frac{\ad^2}{a^2}+ \frac{2 \FDEF \ad
\fd}{F a}+ \frac{\FDEF \fdd}{F}+ \frac{K}{a^2}+ \frac{F_{\f \f}
\fd^2}{F}-
\frac{\fd^2}{4 F}+ \frac{V}{2 F}= 0} \\
~ \\
\disp{ \fdd+ \frac{3 \ad \fd}{a}+ \left( \frac{6 \FDEF F_{\f \f}-
\FDEF}{3 \FDEF^2- F}\right) \frac{\fd^2}{2}+ \frac{2 \FDEF V}{3
\FDEF^2- F}- \frac{F \VDEF}{3 \FDEF^2- F}= 0}
\end{array}
\right. \eeq which do not coincide with the Euler--Lagrange
equations given by (\ref{c2}). Using the first of (\ref{c2}), the
second of (\ref{c2}) can be written as \beq \label{c8}
\begin{array}{l} \disp{\frac{F- 3 \FDEF^2}{F} \fdd+ \frac{3(F- 3
\FDEF^2)}{F} \frac{\ad \fd}{a}+ \left( \frac{\FDEF- 6 F_{\f \f}
\FDEF}{F}\right) \frac{\fd^2}{2}+ \frac{\FDEF \fd^2}{4 F}- \frac{2
\FDEF V}{F}+ \VDEF+}\\~\\
 \disp{+\frac{3 \FDEF \ad^2}{a^2}+ \frac{3 \FDEF
K}{a^2}+ \frac{3 \FDEF^2 \ad \fd}{a}= 0},
\end{array}
\eeq which becomes, taking into account (\ref{c3}) \beq \label{c9}
 \frac{F- 3 \FDEF^2}{F} \fdd+ \frac{3(F- 3 \FDEF^2)}{F} \frac{\ad
\fd}{a}+ \frac{\fd^2}{2 F} \frac{d}{d\f} (F- 3\FDEF^2)+
\frac{\FDEF \fd^2}{4 F}- \frac{2 \FDEF V}{F}+ \VDEF+
\frac{\FDEF}{2 a^3 F} E_t= 0. \eeq Comparing (\ref{c9}) with the
second of (\ref{c7}), we see that they coincide if $F- 3 \FDEF^2
\neq 0$ and $E_t= 0$. The quantity $F- 3 \FDEF^2$ is proportional
to the Hessian determinant of $L_t$ with respect to $(\ad,~\fd)$;
this Hessian has to be different from zero in order to avoid
pathologies in the dynamics \cite{cqg}, while $E_t= 0$ corresponds
to the first order Einstein equation. Clearly, such pathologies
are naturally avoided in the Palatini approach where the
cosmological equations of motion are derived from the field
equations (\ref{40}) and (\ref{42}). It is possible to see more
clearly at the problem of the cosmological conformal equivalence,
formulated in the context of the ``point--like'' Lagrangian, if we
use, as time--coordinate, the conformal time $\eta$, connected to
the cosmic time $t$ by the usual relation \beq \label{c10}
a^2(\eta) d \eta^2= d t^2. \eeq We can see that the use of $\eta$
makes much easier the treatment of all the problems we have
discussed till now. The crucial point is the following: given the
form of the FRW line element expressed in conformal time $\eta$
one does not face the problem of redefining time after performing
a conformal transformation, since in this case, the expansion
parameter appears in front of all the terms of the line element.
From this point of view, the conformal transformation which
connects Einstein and Jordan frame is given by \beq \label{c11}
\left \{
\begin{array}{l}
\abr= \sqrt{-2 F(\f)}\, a \\
~ \\
\disp{ \frac{d\fbr}{d\eta}= \sqrt{\frac{3 \FDEF^2- F}{2 F^2}}\,
\frac{d\f}{d\eta}}
\end{array}
\right. \eeq where $a$, $\f$, $\abr$, $\fbr$ are assumed as
functions of $\eta$.

The Hilbert-Einstein ``point--like'' Lagrangian is given by \beq
\label{c12} \LBR_\eta= -3 \abrpq+ 3 K \abr^2+ \half \abr^2 \fbrpq-
\abr^4 \VBR(\fbr) \eeq in which the prime means the derivative
with respect to $\eta$, and the subscript $\eta$ means that the
time--coordinate considered is the conformal time. Under
transformation (\ref{c11}), it becomes \beq \label{c13}
\begin{array}{ll}
\LBR_\eta & = \disp{ -3 \abrpq+ 3 K \abr^2+ \half \abr^2 \fbrpq-
\abr^4
\VBR(\fbr)=} \\
~ & ~ \\
~ & = \disp{ 6 F(\f) \ap+ 6 \FDEF(\f) a \ap \fp- 6 F(\f) K a^2+
\half a^2 \fpq- a^4 V(\f)= L_\eta}
\end{array}
\eeq which corresponds to the ``point--like'' Lagrangian obtained
from the Lagrangian density in (\ref{s1}) under the hypotheses of
homogeneity and isotropy, using the conformal time as time
coordinate. This means that the Euler--Lagrange equations relative
to (\ref{c12}), which coincides with the second order Einstein
equation and the Klein--Gordon equation in conformal time,
correspond to the Euler--Lagrange equations relative to
(\ref{c13}), under the transformation (\ref{c11}). Moreover, the
energy function $\EBR_\eta$ relative to (\ref{c12}) corresponds to
the energy function $E_\eta$ relative to (\ref{c13}), so that
there is correspondence between the first order Einstein
equations. Furthermore, in order to have full coherence between
the two formulations, it is easy to verify that, both in the
Jordan frame and in the Einstein frame, the Euler--Lagrange
equations, written using the conformal time, correspond to the
Euler--Lagrange equations written using the cosmic time except for
terms in the energy function; for it, one gets the relation \beq
\label{c14} E_\eta= a E_t \eeq which holds in both the frames;
thus the first order Einstein equation is preserved under the
transformation from $\eta$ to $t$ and there is full equivalence
between the two formulations. We want to point out that for the
two Lagrangians $L_\eta$ and $L_t$ the same relation as
(\ref{c14}) holds. On the other hand, such results naturally hold
if one takes into account the relation (\ref{bimetri1}) derived
from the second Palatini equation (\ref{40}).

When ordinary matter is present the standard Einstein
(cosmological) ``point--like'' Lagrangian  is  \beq \label{c15}
\LBR_{tot}= \LBR_\tbr+ \LBR_{mat}, \eeq in which $\LBR_\tbr$ is
given by (\ref{c5}) and $\LBR_{mat}$ is the Lagrangian relative to
perfect fluid matter. Using the contracted Bianchi identity, it
can be seen that $\LBR_{mat}$ can be written as \cite{matter} \beq
\label{c16} \LBR_{mat}= -D \abr^{3(1- \gam)}, \eeq where $D$ is
connected to the total amount of matter. In writing (\ref{c15})
and (\ref{c16}) we have chosen the cosmic time as
time--coordinate. Under the transformation (\ref{c4}) we have,
besides relation (\ref{c5}), that (\ref{c16}) corresponds to \beq
\label{33} \LBR_{mat}= (\sqrt{-2 F})^{3(1- \gam)} L_{mat}, \eeq
where, analogously to (\ref{c16}) \beq \label{c18} L_{mat}= D
a^{3(1- \gam)}. \eeq Then we have that, using (\ref{c4}),
(\ref{c15}), it  becomes \beq \label{c19} \frac{1}{\sqrt{-2 F}}
L_{tot} =\frac{1}{\sqrt{-2 F}} [L_t+ (\sqrt{-2 F})^{(4- 3 \gam)}
L_{mat}] \eeq in which we have defined the total ``point--like''
Lagrangian after the conformal transformation as \beq \label{c20}
L_{tot}= L_t+ (\sqrt{-2 F})^{(4- 3 \gam)} L_{mat}, \eeq (cfr.
(\ref{c5})); the transformation of $\LBR_{tot}$ under (\ref{c4})
has to be written following the expression (\ref{c19}) and
consequently the ``point--like'' Lagrangian $L_{tot}$ has to be
defined as in (\ref{c20}).

Summarizing, the perfect fluid-matter, which minimally interact in
the Jordan frame, results non-minimaly interacting in the
conformally transformed Einstein frame unless $\gam= \frac{4}{3}$
(radiation), since the standard matter Lagrangian term is coupled
with the scalar field in a way which depends on the coupling $F$.
Such a coupling between the matter and the scalar field is an
effect of the transformation, therefore depending on the coupling.
Also this interaction which emerges passing from the Jordan frame
to the Einstein frame, is immediately recovered considering the
Palatini structural equation (\ref{stru1}) and follows directly
from (\ref{47}) which express the relation between the different
metrics (and consequently between the two frames).

\subsection{Some relevant examples}

The exact identification of the frame is crucial when the
solutions are matched with data. We are going to give some
examples where the nature of solutions drastically changes
considering the Einstein frame or the Jordan frame without taking
into account the problem of transformations of physical quantities
between them. The ambiguity is removed in the Palatini approach
since, due to the intrinsic bi-metric structure, the two frames
are given together by the same dynamics.

{\bf i)} Let us consider a model in the Einstein frame with a
scalar field, a constant potential and zero curvature. The
Lagrangian is given by \beq \label{e1} \LBR_\tbr= -3 \abr \abrd^2+
\half \abr^3 \fbrd^2- \abr^3 \LAM; \eeq the Euler--Lagrange
equations and the  energy  condition are \beq \label{e2} \left \{
\begin{array}{l}
\disp{ \frac{2 \abrdd}{\abr}+ \frac{\abrd^2}{\abr^2}+ \half
\fbrd^2-
\LAM= 0} \\
~ \\
\disp{ \fbrdd+ \frac{3 \abrd \fbrd}{\abr}= 0}.
\end{array}
\right. \eeq \beq \label{e3} \frac{\abrd^2}{\abr^2}- \frac{1}{3}
\left( \half \fbrd^2+ \LAM\right)= 0. \eeq The system can be
easily solved giving the solution \beq \label{e4} \left \{
\begin{array}{l}
\abr= \left[ c_1 e^{\sqrt{3 \LAM} \, \tbr}- \disp{
\frac{\fbrd_0^2}{8
\LAM c_1^2}} e^{-\sqrt{3 \LAM} \, \tbr}\right]^{\frac{1}{3}} \\
~ \\
\fbr= \fbr_0+ \disp{ \sqrt{\frac{2}{3}}}\, \ln{\frac{1- \disp{
\frac{\fbrd_0}{2 c_1 \sqrt{2 \LAM}}} e^{-\sqrt{3 \LAM} \,
\tbr}}{1+ \disp{ \frac{\fbrd_0}{2 c_1 \sqrt{2 \LAM}}} e^{-\sqrt{3
\LAM} \, \tbr}}}
\end{array}
\right. \eeq Three integration constants  appear in the solution,
since Eq. (\ref{e3}) corresponds to a constraint on the value of
the first integral $\EBR_\tbr$. We have that, in the limit of
$\tbr \rightarrow +\infty$, the behavior of $\abr$ is exponential
with characteristic time given by $\sqrt{\frac{\LAM}{3}}$, as we
would expect, and $\fbr$ goes to a constant. Looking at the second
of (\ref{s10}), we have that such a model in the Einstein frame
corresponds, in the Jordan frame, to the class of models with
(arbitrarily given) coupling $F$ and potential $V$ connected by
the relation \beq \label{e5} \frac{V}{4 F^2}= \LAM, \eeq the
solution of which can be obtained from (\ref{e4}) via the
transformation (\ref{c4}). We can thus fix the potential $V$ and
obtain, from (\ref{e5}), the corresponding coupling. This can be
used as a method to find the solutions of non-minimally coupled
models with given potentials, the coupling being determined by
(\ref{e5}). In other words, a single model in the Einstein frame
corresponds to a family of models in the Jordan frame, but giving
``a priori" the bi-metric structure of the theory by the Palatini
approach, the model is only one. As an example, let us take into
account the case \beq \label{e6} V= \lam \f^4,~~~ \lam> 0 \eeq
which correspond to a ``chaotic inflationary'' potential
\cite{infl}. The corresponding coupling is quadratic in $\f$ \beq
\label{e7} F= k_0 \f^2 \eeq in which \beq \label{e8} k_0= -\half
\sqrt{\frac{\lam}{\LAM}}. \eeq Substituting (\ref{e4}) into
(\ref{c4}), we get \beq \label{e9} \left \{
\begin{array}{l}
\disp{ a= \frac{\abr}{\f \sqrt{-2 k_0}}} \\
~ \\
\disp{ d\f= \f \sqrt{\frac{2 k_0}{12 k_0 -1}}\, d\fbr} \\
~ \\
\disp{ dt= \frac{d\tbr}{\f \sqrt{-2 k_0}}}.
\end{array}
\right. \eeq As we see from these relations, it has to be $k_0<
0$. Integrating the second of (\ref{c9}), we have the conformal
relation between the scalar fields, \ie $\f$ in terms of $\fbr$
\beq \label{e10} \f= \al_0 e^{\sqrt{\frac{2 k_0}{12 k_0- 1}} \,
\fbr}. \eeq Substituting (\ref{e10}) in the first of (\ref{e9})
and taking into account the second of (\ref{e4}), we have the
solutions $a$ and $\f$ as functions of $\tbr$ \beq \label{e11}
\left \{
\begin{array}{l}
\f= \f_0 \left[ \frac{\disp{ 1- \frac{\fbrd_0}{2 c_1 \sqrt{2
\LAM}}} e^{-\sqrt{3 \LAM} \, \tbr}}{\disp{ 1+ \frac{\fbrd_0}{2 c_1
\sqrt{2 \LAM}}} e^{-\sqrt{3 \LAM} \, \tbr}}\right]^{\sqrt{\frac{4
k_0}{3(12 k_0-
1)}}} \\
~ \\
a= \disp{ \frac{1}{\f_0 \sqrt{-2 k_0}}} \left[ c_1 e^{\sqrt{3
\LAM} \, \tbr}- \disp{ \frac{\fbrd_0^2}{8 \LAM c_1^2}} e^{-\sqrt{3
\LAM} \, \tbr}\right]^{\frac{1}{3}} \left[ \frac{\disp{ 1+
\frac{\fbrd_0}{2 c_1 \sqrt{2 \LAM}}} e^{-\sqrt{3 \LAM} \,
\tbr}}{\disp{ 1- \frac{\fbrd_0}{2 c_1 \sqrt{2 \LAM}}} e^{-\sqrt{3
\LAM} \, \tbr}}\right]^{\sqrt{\frac{4 k_0}{3(12 k_0- 1)}}}
\end{array}
\right. \eeq in which $\f_0= \al_0 e^{\sqrt{\frac{2 k_0}{12 k_0-
1}} \, \fbr_0}$. Substituting (\ref{e10}) in the third of
(\ref{e9}), taking into account (\ref{e4}), we get \beq
\label{e12} dt= \frac{d\tbr}{\f_0 \sqrt{-2 k_0}} \left[ \frac{1+
\disp{ \frac{\fbrd_0}{2 c_1 \sqrt{2 \LAM}}} e^{-\sqrt{3 \LAM} \,
\tbr}}{1- \disp{ \frac{\fbrd_0}{2 c_1 \sqrt{2 \LAM}}} e^{-\sqrt{3
\LAM} \, \tbr}}\right]^{\sqrt{\frac{4 k_0}{3(12 k_0- 1)}}}. \eeq
We obtain $\tbr$ as a function of $t$ integrating (\ref{e12}) and
then considering the inverse function; Eq.(\ref{e12}) could be
easily integrated if the exponent $\sqrt{\frac{4 k_0}{3(12 k_0-
1)}}$ would be equal to $\pm 1$, but this corresponds to a value
of $k_0= \frac{3}{32}$ which is positive and thus it turns out to
be not physically acceptable. In general, (\ref{e12}) is not of
easy solution. We can analyze its asymptotic behavior, obtaining
\beq \label{e13} \frac{dt}{d\tbr} \stackrel{\tbr \rightarrow
+\infty}{\rightarrow} \frac{1}{\f_0 \sqrt{-2 k_0}} \eeq that is,
asymptotically, \beq \label{e14} t- t_0 \simeq \frac{\tbr}{\f_0
\sqrt{-2 k_0}}. \eeq Substituting (\ref{e14}) in the asymptotic
expression of (\ref{e11}), we obtain the asymptotic behavior of
the solutions (since from (\ref{e13}) one has $t \stackrel{\tbr
\rightarrow +\infty}{\rightarrow} +\infty$) \beq \label{e15} \left
\{
\begin{array}{l}
a \simeq \disp{ \frac{c_1^{1/3}}{\f_0 \sqrt{-2 k_0}}} e^{\f_0
\sqrt{\frac{-2
\LAM k_0}{3}} \, (t- t_0)} \\
~ \\
\f \simeq \f_0.
\end{array}
\right. \eeq Thus we have that, asymptotically, $a(t)$ is
exponential, and $\f(t)$ is constant; the coupling $F$ is
asymptotically constant too, so that, fixing the arbitrary
constant of integration to obtain the finite transformation of
$\abr$, $\fbr$ (that is, fixing the units, see \cite{sokolowski}),
once $k_0$ is fixed, it is possible to recover asymptotically the
Einstein gravity from the Jordan frame.

As a remark we would like to notice that the asymptotic expression
(\ref{e15}) of $a(t)$ and $\f(t)$ are solutions of the Einstein
equations and Klein--Gordon equation with zero curvature and $F$
and $V$ given by (\ref{e6}), (\ref{e7}). They have not been
obtained as solutions of the asymptotic limits of these equations.
It means then that they are, in any case, particular solutions of
the given non-minimally coupled model.

{\bf ii)} Another interesting case is the Ginzburg--Landau
potential \beq \label{e16} V= \lam (\f^2- \mu^2)^2,~~~ \lam> 0.
\eeq The corresponding coupling is given by \beq \label{e17} F=
k_0 (\f^2- \mu^2), \eeq in which $k_0$ is given by (\ref{e8}) when
$\f^2> \mu^2$ while is given by (\ref{e8}) with opposite sign when
$\f^2< \mu^2$, in order to have $F< 0$. With this coupling, the
corresponding conformal transformation turns out to be singular
for $\f^2= \mu^2$, thus with this method it is not possible to
solve this model for $\f$ equal to the Ginzburg--Landau mass
$\mu$. The explicit function $\f= \f(\fbr)$ is obtained inverting
the integral \beq \label{e18} \fbr- \fbr_0= \int \frac{[3
\sqrt{\frac{\lam}{\LAM}}\f^2+ \half (\f^2-
\mu^2)]^\half}{[\frac{\lam}{4 \LAM}]^{\frac{1}{4}} (\f^2-
\mu^2)}\, d\f; \eeq and it is possible to carry  analogous
considerations as in the previous case, concluding that
asymptotically the behavior of $a(t)$ is exponential and that of
$\f(t)$ is constant.

{\bf iii)} Another interesting case is  \beq \label{e19} V= \lam
\f^2,~~ \lam> 0;~~~ F= k_0 \f^2,~~ k_0< 0 \eeq in the Jordan
frame.  The coupling is the same as in  (\ref{e7})) and the
conformal transformation is given by (\ref{e9}). To obtain the
corresponding potential in the Einstein frame we have to
substitute (\ref{e10}) in the relation \beq \label{e20}
\VBR(\fbr)= \frac{\lam}{4 k_0^2 \f^2(\fbr)}, \eeq that is \beq
\label{e21} \VBR(\fbr)= \frac{\lam}{4 k_0^2 \f^2_0} e^{-2
\sqrt{\frac{2 k_0}{12 k_0- 1}} \, \fbr}, \eeq which gives, in the
Einstein frame,  power--law solutions \cite{lucchin-mat,barrow}. A
general remark concerns the relation between the Hubble parameter
in the Einstein and in the Jordan frame. It is \beq \label{e22}
\bar{H}  = \frac{\abrd}{\abr}= \frac{1}{(-2 F)} \left(
-\frac{\FD}{\sqrt{-2 F}}+ \sqrt{-2 F} \frac{\ad}{a}\right)
=\frac{\FD}{2 F \sqrt{-2 F}}+ \frac{H}{\sqrt{-2 F}} , \eeq in
which we have used the relations (\ref{c4}). Relation (\ref{e22})
is useful to study the asymptotic behavior of the Hubble
parameter: if we require an asymptotic de Sitter--behavior in both
the Einstein and Jordan frame (for example, in order to reproduce
quintessential accelerated behavior),  we have to require $\bar{H}
\stackrel{\tbr \rightarrow +\infty}{\rightarrow} \bar{C}$ and $H
\stackrel{t \rightarrow +\infty}{\rightarrow} C$ where $\bar{C}$
and $C$ are constants, from (\ref{e22}), we obtain a differential
equation for the coupling $F$ as a function of $t$ ($t>> 0$),
given by \beq \label{e23} \FD+ 2 C F- 2 \bar{C} F \sqrt{-2 F}= 0.
\eeq Its solution is \beq \label{e24} F= -\frac{C^2}{2 \bar{C}^2}
\left[ \frac{1}{1- F_0 e^{C t}}+ 1\right]^2, \eeq in which $F_0$
is the integration constant; this is the time--behavior that $F$
has to assume on the solution $\f(t)$, in order to have a de
Sitter asymptotical accelerated behavior in both frames.  It easy
to verify that both the couplings in the examples i) and ii)
satisfy (\ref{e24}) asymptotically.

\section{ Conclusions}
The decennial debate about the physical relevance of conformal
transformations can be enlightened by taking into account the
Palatini approach to the Extended Theories of gravity. In this
framework, the conformal transformation is not only a mathematical
tool capable of disentangling matter from gravitational degrees of
freedom, but it is related to the bi-metric structure of spacetime
where chronological structure and geodesic structure are, a
priori, independent. While in Hilbert-Einstein gravity the affine
connections can be assumed in any case Levi-Civita, this is not
true in the Palatini approach, being the fields $g$ and $\Gamma$
independent. Due to this fact, the ambiguities to work out a given
theory in the Einstein or in the Jordan frame are removed since
the Palatini field equations and, first of all the structural
equation of spacetime, give, at the same time, information on both
frames. In other words, discussing if ``Jordan" or ``Einstein" is
the true physical frame is a nonsense in the Palatini approach.

 In this paper, we have taken into account scalar-tensor, higher-order
and higher-order-scalar-tensor theories of gravity showing how the
Palatini field equations furnish the conformal structure.
Furthermore, we have shown that higher-order and scalar-tensor
theories can be dealt under the same standard: in particular,
scalar-tensor (second-order) theories can be recovered in the
limit of small curvatures.

These results become crucial in cosmology since, by them,  it is
possible to show that solutions taken into account as different
ones are the same in the Palatini approach. For example, the
recently observed acceleration of the Hubble fluid
\cite{snIa,debernardis,Perlmu,Riess,Verde} is an evidence that
some form of ``dark energy" should be present in the cosmic
dynamics. Despite of this general result, such an accelerated
dynamics can be achieved in several ways (cosmological constant
\cite{cosmconst}, scalar fields dynamics \cite{steinhardt},
curvature quintessence \cite{carrol2}) but no definite answer, up
to now, has been given about its nature. In what we have
discussed, we have shown that a cosmic dynamics ruled by the
cosmological constant in the Einstein frame becomes ruled by a
non-minimally coupled, self-interacting scalar field (evolving in
time) in the Jordan frame. Consequently, matching the data against
a solution in the Einstein frame or in the Jordan frame could lead
to highly misleading results and interpretation. The shortcoming
is completely overcome in the Palatini approach which furnishes,
at the same time, dynamics and conformal structure of the given
ETG avoiding such ambiguities. These considerations have to be
further developed considering concretely the matching with the
data.

{\bf Acknowledgements}\\
\noindent We are  grateful  for  useful discussions and
suggestions to V. Cardone, S. Carloni and A. Troisi.
\\
This work is partially supported (G.A. and M.F.)  by GNFM--INdAM research project ``\emph{Metodi geometrici
in meccanica classica, teoria dei campi e termodinamica}'' and by MIUR: PRIN 2003 on
``\emph{Conservation laws and thermodynamics in continuum mechanics and field theories}''.
G.A. is supported by the I.N.d.A.M. grant: ``Assegno di collaborazione ad attivit\'a di ricerca a.a. 2002-2003".


\end{document}